\def\ln{\ell{n}}
\documentstyle[12pt]{article}

 {\parindent 0.5cm \textwidth 15.0cm \textheight
23.5cm \hoffset=-0.8cm \voffset=-3cm
  \let\LARGE=\large
 \let\large=\normalsize
\begin{document}
\begin{titlepage} \vspace{0.2in} \begin{flushright}
MITH-93/6 \\ \end{flushright} \vspace*{1.5cm}
\begin{center} {\LARGE \bf  Mass Relation Between Top and Bottom Quarks
\\} \vspace*{0.8cm}
{\bf Giuliano Preparata$^{a,b}$ and She-Sheng Xue$^{a,\dagger}$\\}
a) INFN, Section of Milan, Via Celoria 16, Milan, Italy\\
b) Physics Department, University of Milan, Italy\\
\vspace*{1.8cm}
{\bf   Abstract  \\ } \end{center} \indent

In the framework of the recently proposed electroweak theory on a Planck
lattice, we are able to solve approximately the lattice Dyson equation for
the fermion self-energy functions, and obtain the ratio between the masses of
the $t-$ and $b-$ quarks in terms of the electroweak coupling constants. The
predicted top mass agrees with recent determinations from electroweak
observables.
\vfill \begin{flushleft} 1st February, 1993   \vspace*{3cm} \\
\noindent{\rule[-.3cm]{5cm}{.02cm}} \\
\vspace*{0.2cm} \hspace*{0.5cm} ${}^{\dagger}$
E-mail address: xue@milano.infn.it\end{flushleft} \end{titlepage}

In recent papers \cite{xue91}\cite{xue92}\cite{xue93} we have proposed to
incorporate in the electroweak theory the possible effects of violent gravity
quantum fluctuations at the Planck scale ($a_p \simeq 10^{-33} cm$) by means of
a (random) space-time lattice structure, whose lattice constant is just $a_p$.
Remarking that such a lattice structure, a kind of ``worm-hole" condensation in
the ground state of the quantum gravitational field, is not a new proposal, but
it is suggested by some analysis of the small scale quantum fluctuations of
gravity \cite{planck}. We noticed that the well known ``no-go" theorem of
Nielsen and Ninomiya \cite{nogo} would not allow a simple transcription of the
electroweak lagrangian on such lattice -~to be called Planck lattice~-, and
argued for the necessity to extend the usual electroweak Lagrangian of the
Standard Model to include effective gauged Nambu-Jona Lasinio (NJL)
\cite{nambu} types of interactions, quadrilinear in the Fermi-field. In this
way the gauge principle could be obeyed while avoiding the difficulties of the
``no-go" theorem \cite{xue91}. In addition, even though the NJL-terms are at
this time simply added for consistency, it is however possible to envisage
their origin as ``effective" interactions induced by Quantum Gravity (QG), thus
tying finally together in a fundamental way the physics of the Standard Model
to QG.

As a first step in the development of our program we neglected the
usual gauge-field interactions, and analyzed the solutions of the Dyson
equations involving the NJL interactions only with the following
results\cite{xue92}:
\begin{enumerate}
\item
The fundamental chiral symmetry of the full Lagrangian is
spontaneously broken, and as a consequence \underbar{only one quark family},
which is identified with the top (t) quark and bottom (b) quark doublet,
acquires a mass, that within the mentioned approximation is the same for both,
i.e. $m_t = m_b$;

\item
The consistent solution of the gap equations produces also non-zero
Wilson-type parameters $r_q \simeq 0.3$ and mass counterterms for each quark,
thus solving ``in
practice" the difficulties of the ``no-go" theorem, by removing through the
Wilson-mechanism the unobserved ``doublers";

\item
The composite Goldstone particles carry the quantum numbers of the gauge
bosons $W^{\pm}$ and $Z^0$, and should end up as the longitudinal modes of
these gauge bosons. The mechanism by which the Goldstone
particles get ``eaten up"
by the gauge bosons, that in so doing become massive, has been discussed in the
NJL-context in Refs.\cite{bardeen}.
The composite scalar, on the other hand,
that in the continuum theory replaces the Higgs meson\cite{higgs}, is seen to
acquire a mass of the
order of the Planck mass ($m_p \simeq 10^{19} \; GeV$), thus becoming
unobservable.

\item
The effective action that is left after such massive rearrangement of the
vacuum is ($a=a_p$)
\begin{eqnarray}
S & = & S_G + S_D + \sum_{xF} \bar \psi^F (x) m_F \psi (x)\label{action}\\
&& - {1 \over 2a} \sum_{Fx \mu} \bigl[ \bar \psi^F (x) \left(L^F_\mu(x) +
R^F_\mu(x )\right) r_F U^F_\mu (x) \psi^F (x + a_\mu) + {\rm h.c.} \bigr]
+\cdot\cdot\cdot,
\nonumber
\end{eqnarray}
where $S_G$ is the usual Wilson gauge-action, $S_D$ the usual Dirac action,
$F=l (q)$ denotes the lepton (quark) sector, $m_F, r_F$ are matrices in
flavour and weak isospin space, and ``$\cdot\cdot\cdot$'' denotes the necessary
counterterms. Finally
\begin{equation}
L^F_\mu(x) = U^L_\mu(x) V^{Y^F_L}_\mu(x);\;\;G^R_\mu(x) =
\pmatrix{
V^{Y^F_{R_1}}_\mu (x) &         0           \cr
0               &   V_\mu^{Y^F_{R_2}} (x)   \cr} ,
\label{l}
\end{equation}
and
\begin{equation}
U^c_\mu(x) \in SU_c(3), \;\;U^L_\mu(x) \in SU_L(2) \; {\rm and }\; V_\mu(x)
\in U_Y(1).
\label{gauge}
\end{equation}
\end{enumerate}

In this note we go one step forward and study the Dyson equations for the
massive quark doublet ($t,b)$ including the interactions with the gauge fields
$W^{\pm}$, $Z_0$ and $\gamma$(photon) based on the action (\ref{action}). Our
aim is, of course, to see whether these latter interactions are capable to lift
the identity of $m_t$ and $m_b$, which is experimentally known to be badly
violated.

We take into consideration of the NJL interaction and gauge interactions and
the Dyson equations will thus have the structure depicted diagrammatically in
Fig.1. The Landau mean-field and the large-$N_c$ approach have been adopted.
For external momenta $p_\mu a \ll 1$, we divide the integration domain over the
variable $q_\mu$ in two regions: the ``continuum" region: $0 \leq |q_\mu a|
\leq \epsilon$ and the ``lattice" region: $\epsilon \leq |q_\mu a| \leq \pi$,
where $|ap_\mu| \ll \epsilon \ll \pi$. With this separation we can write the
Dyson equation as:
\begin{eqnarray}
\Sigma^{t,b}_c (p) & = & m_t (NJL)^{t,b} + \left[ C_{\gamma}(p) + C_{QCD}(p) +
C_{Z_0} (p)\right]^{t,b}\nonumber\\
&& + \left[L_{\gamma}(p) + L_{QCD} (p) + L_{Z_0} (p) + L_W (p) \right]^{t,b},
\label{t}
\end{eqnarray}
where $\sum^{t,b}_c$ denotes the self-energy function of $t-$ and $b-$ quarks,
respectively in the ``continuum region", and
\begin{eqnarray}
(NJL)^{t,b} &=& 2g_1 \int ^\pi _{-\pi} {d^4q \over (2 \pi)^4} \; {1 \over
\sin^2
q_\mu + [m_{t,b} a + rw(q)]^2},
\label{njl}\\
C^{t,b}_g (p) &=& {1\over 4 \pi^2} \int_{\Lambda_p} \lambda^c_g (q) d^4q {1
\over (p-q)^2 + m^2_g} \; {\sum^{t,b}_c (q) \over q^2 + m^2_{t,b}},
\label{c}
\end{eqnarray}
where the subscript $``g"$ denotes the relevant gauge-interaction, with the
appropriate running gauge-coupling $\lambda^c_g(q) = {3 \over \pi} \; {g^2_c(q)
\over 4 \pi}$ in the ``continuum region"; $g_1$ is the dimensionless coupling
introduced in\cite{xue92}, and $w(q) = \sum_\mu (1 - \cos q_\mu a)$. Note that
no $W^{\pm}$-term exists in Eq.~(\ref{t}) in the ``continuum" region due to the
unique chirality of W-exchange in this region. Note also that the integrals
over the ``continuum" region run up to $\Lambda_p$ and not to
$\Lambda=\epsilon\Lambda_p$ for, as shown in\cite{xue93}, the $\ln
\epsilon$-term contained in $L_g$ (with the exception of $L_W$) has been added
to the ``continuum" region integral, thus leading to an overall
$\epsilon$-independence of the terms appearing in Eq.~(\ref{t}), as it should
happen. The $\epsilon$-independent parts $\bar L_g$ of $L_g$ can in general be
determined by numerical fitting:
\begin{equation}
\lambda^L_g m_{t,b} (\bar L^{t,b}_g (r) + \; \theta \; \ln \epsilon ) =  -
\lambda^L_g \int_{[\epsilon, \pi]} {d^4l \over (2 \pi)^4} \cdot
{\sum^{t,b}_c (l) \over 4 \sin^2 {l_\mu \over 2}}
  \cdot {- \cos^2{ l_\mu\over 2} + r^2 x \sin^2 {l_\mu\over2} \over
\sin^2 l_{\mu} + (rw(l))^2 },
\label{fit}
\end{equation}
where the approximation $\sum^{t,b}_c (l) \simeq m_{t,b}$ is made for
momenta $l_\mu\in [\epsilon, \pi]$.
$\bar L^{t,b}_g(r)$ is plotted as a function of the Wilson parameter $r$ in
Fig.2. In Eq.~(\ref{fit}) $x = 1$ for $L_{\gamma}$ and $L_{QCD}$, and
\begin{equation}
x = \left\{
\matrix{
{3 \over 2} {1 \over \sin^2 \theta^L_w ({1 \over 3} \sin^2 \theta^L_w +
  \cos^2 \theta^L_w ) } & {\rm for \; the \; {\it b}-quark} \cr
& \cr
{3 \over 4} {1 \over \sin^2 \theta^L_w (\cos^2 \theta^L_w - {1 \over 3}
\sin^2 \theta^L_w ) } & {\rm for \; the \; {\it t}-quark}
} \right.
\label{x}
\end{equation}
for the $L_{Z_0}$ contribution. The $W^{\pm}$-contribution $L_W$ is given by
\begin{equation}
L^{t,b}_W = - \lambda^L_W \int^\pi_{- \pi} {d^4l \over 16 \pi^2} \; {r^2 \over
4} \; {\sum^{b,t}_c (l) \over \sin^2 l_\mu + [rw(l)]^2},
\label{w}
\end{equation}
and will play a very important r\^ole in the $t-b$ mass splitting. The gauge
coupling $\lambda^L_g={3\over\pi}{g^2_L\over 4\pi}$ and
$\lambda^L_W={3\over\pi} {\alpha^L_{QED}\over \sin^2\theta^L_w}
(\sin^2\theta^L_w\simeq 0.5$), the ``lattice'' gauge couplings, are accordingly
determined by the `` bare '' coupling constants. As for the ``lattice" region
terms, denoted in Eq.~(\ref{t}) by $L_i(p)$, we remark that the terms that
diverge like ${1 \over a}$, contained in $L^{t,b}_g$, can all be consistently
cancelled by mass counterterms\cite{xue92}.

The complicated system (\ref{t}) can now be enormously simplified by
setting $p_{\mu} a \simeq 0$ and by neglecting all momentum
dependences in $\sum_c^{t,b} (l)$, thus replacing them by $m_t$ and
$m_b$ respectively. We note right away that $m_t = m_b$ is no more a
solution of the ensuing system, indeed the simplified Eq.~(\ref{t})
for the $t$-quark becomes:
\begin{equation}
m_t [1- (NJL) - m_t\sum_g ( \bar C_g^t + \bar L^t_g)] = m_b\bar L_W,
\label{mt}
\end{equation}
where $\bar C$ and $\bar L$ are the contributions discussed above divided
by the ``mean-field" value $\sum^{t,b}_c (q) = m_{t,b}$. Analogously for
the $b$-quark we have
\begin{equation}
m_b [1 - (NJL) - \sum_g (\bar C^b_g + \bar L^b_g) ] = m_t \bar L_W,
\label{b}
\end{equation}
where due to $m_t a \simeq m_b a \simeq 0$, $ (NJL)$, $\bar L_{QCD}$
and $\bar L_W$ are the same for both quarks.
Our philosophy is that the world is so constructed as to yield small masses in
a theory that starts out with only one mass scale, the Planck scale. In order
for
this to arise in the above two equations (\ref{mt}) and (\ref{b}), one must
``tune'' the $NJL$-term (the coupling $g_1$ in (\ref{njl})) so that the LHS
of (\ref{mt}) is a small number. Now if we make the above ``tuning'', the
analogous term in (\ref{b}) cannot be so small due to
$\bar C_g^b\not=\bar C_g^t$ and $\bar L_{Z_0}^b\not=\bar L_{Z_0}^t$.
It is clear that our ``tuning'' does not allow us to determine both
$m_t$ and $m_b$, but only their ratio ${m_t\over m_b}$. Thus from
Eqs.~(\ref{mt}) and (\ref{b}) we get trivially
\begin{equation}
\bar L_W m^2_t - 2 \Delta m_b m_t - \bar L_W m^2_b = 0,
\label{y}
\end{equation}
where
\begin{equation}
2 \Delta = (\bar C^t_{\gamma} - \bar C^b_{\gamma})+(\bar C^t_{Z_0} - \bar
C^b_{Z_0}) + (\bar C^t _{QCD} - \bar C^b _{QCD}),
\label{z}
\end{equation}
note that $\bar L_{Z_0}^{t,b}$ is negligible with respect to
$\bar C_g^{t,b}$. Furthermore, the physical $Z_0$ boson and gluon masses being
very heavy, $\bar C^t_{QCD} - \bar C^b_{QCD}$ and $\bar C^t_{Z_0} - \bar
C^b_{Z_0}$ can also be neglected, we can approximate 2$\Delta$ as
\begin{equation}
2\Delta \simeq (\bar C^t_{\gamma} - \bar C^b_{\gamma}).
\label{ap}
\end{equation}
In Eq.~(\ref{c}), the running coupling constant $\lambda^c_{QED} (q^2)$ is
introduced by
\begin{equation}
\alpha^c_{QED} (q^2) = \left( 1 - {\alpha ^c_{QED} (q^2) \over3 \pi} \;
\sum_f \; Q^2_f \; \ln \; {q^2 \over m^2_f}\right) \alpha^L_{QED},
\label{j}
\end{equation}
where all fermion loop contributions have been taken into account.
Thus form Eq.~(\ref{c}) we obtain
\begin{eqnarray}
2 \Delta & \simeq & - 0.0025 \; {\pi \alpha^L_{QED} \over 24} + 0.0025 \;
{3 \alpha^L_{QED} \over 8 \pi} \; \bigg[ {4 \over 9} \left( \ln \;
{\Lambda^2_p \over
m^2_t} \right)^2\nonumber\\
&& - {1 \over 9} \left(\ln \; {\Lambda^2_p \over m^2_b}\right)^2 \bigg] +
0.644 \left( {3 \alpha^L_{QED} \over 2 \pi} \right) \left[ {4 \over 9} \;
\ln  \; {\Lambda _p \over m_t} - {1 \over 9} \; \ln \; {\Lambda_p
\over m_b} \right].
\label{k}
\end{eqnarray}
As for $\bar L_W$ we have
\begin{equation}
\bar L_W = 6 \pi \alpha_L \; {r^2 \over 4} \; G(r);\hskip2cm
G(r) = \int^{\pi}_{- \pi} d^4l \; {1 \over \sin^2 l_{\mu} + r^2W(l)^2},
\label{f}
\end{equation}
where $G(r)$ is plotted in Fig.3. For $r =r_m \sim 0.3$, the ground
state\cite{xue92}, $G(r_m) \simeq 0.325$.
We are now in a position to solve Eq.~(\ref{y})
\begin{equation}
m_t =m_b {\Delta + \sqrt{\Delta^2 + \bar L^2_W } \over \bar L_W}
\label{r}
\end{equation}
By substituting Eqs.~(\ref{k}) and (\ref{f}) into Eq.~(\ref{r}), we find
that a consistent solution to (\ref{r}) is
\begin{equation}
|m_t| \simeq 30|m_b|.
\label{v}
\end{equation}
Setting $m_b = 4.7 \; GeV$, as given by a theoretical analysis of the $q \bar
q$-spectrum \cite{qq}, we predict
\begin{equation}
m_t \cong 145 \; GeV,
\label{re}
\end{equation}
which appears to agree with indirect determinations from electroweak parameters
\cite{exp}. Let us point out that we achieve here for the first time
the goal of relating the very different masses of the top and bottom quarks to
their different charges, to the Planck length, and lattice effects $\bar
L_w$, even though our analysis of fermion masses is still at a preliminary
stage
and the main uncertainty comes from our approximate way of determining the
Wilson parameter $r_q$.

The fact that we have been able to obtain a relation between the
masses of the two quarks that become massive from the spontaneous
breaking of the chiral symmetry, that occurs in the NJL-extensions of
the electroweak theory on the Planck lattice, appears to us as a
rather pleasing signal of the physical relevance of the ideas that we
have been developing recently. To summarize, the picture that is
emerging from our work is that
\begin{enumerate}
\item
the standard low-energy electroweak theory is the low-energy (much smaller than
the Planck mass) approximation of a chirally symmetric theory on a Planck
lattice;

\item
mass gets generated spontaneously, and to a first stage a quark doublet becomes
massive, together with the gauge bosons $W^{\pm}$ and $Z^0$;

\item
the generation of the gauge bosons' masses does not produce any additional
massive particle, like the Higgs boson, for on a Planck lattice scalar states
get lifted to the Planck mass \cite{xue92};

\item
by taking due account of all the gauge interactions, in a reliable
approximation to the gap equations, we are able to determine the mass
ratio of the two quarks that become massive, that turns out to be in agreement
with the indirect information that is now available.
\end{enumerate}

So far so good; but what lies ahead? Clearly we must study
the relation between the $W^{\pm}$ and $Z^0$ masses and the $t$-quark
mass. When this is understood we
will be able to turn our attention to the possible origins of the
masses of the other fermions, as well of the CKM mass-matrix; and the
strategy followed in this paper appears rather promising.
But these are only speculations: thus it is time to stop.

One of us - (S.-S.~Xue) - wishes to thank Prof.~Hai-Yang~Cheng, Hai-Li~Yu and
High-Energy~Physics~Group in the Institute of Physics, Academia~Sinica for the
kind hospitality extended to him at their Institutions, where part of this work
was done.

\newpage  \pagestyle{empty}
\begin{center} \section*{Figure Captions} \end{center}

\vspace*{1cm}

\noindent {\bf Figure 1}: \hspace*{0.5cm}
The Dyson equations for top and bottom quarks.

\noindent {\bf Figure 2}: \hspace*{0.5cm}
The function $\bar L(r)$ in terms of $r_q$ for $x=1$.

\noindent {\bf Figure 3}: \hspace*{0.5cm}
The function $G(r)$ in terms of $r_q$.

}
\end{document}